# Superior Structural, Elastic and Electronic Properties of 2D Titanium Nitride MXenes Over Carbide MXenes: A Comprehensive First Principles Study

Ning Zhang[1], Yu Hong[1], Sanaz Yazdanparast[1], and Mohsen Asle Zaeem[1,2*]

[1] Department of Materials Science and Engineering, Missouri University of Science and Technology, Rolla, MO 65409, USA

[2] Department of Mechanical Engineering, Colorado School of Mines, Golden, CO 40801, USA

**Abstract**

The structural, elastic and electronic properties of two-dimensional (2D) titanium carbide/nitride based pristine ($Ti_{n+1}C_n$/$Ti_{n+1}N_n$) and functionalized MXenes ($Ti_{n+1}C_nT_2$/$Ti_{n+1}N_nT_2$, T stands for the terminal groups: –F, –O and –OH, n = 1, 2, 3) are investigated by density functional theory calculations. Carbide-based MXenes possess larger lattice constants and monolayer thicknesses than nitride-based MXenes. The in-plane Young's moduli of $Ti_{n+1}N_n$ are larger than those of $Ti_{n+1}C_n$, whereas in both systems they decrease with the increase of the monolayer thickness. Cohesive energy calculations indicate that MXenes with a larger monolayer thickness have a better structural stability. Adsorption energy calculations imply that $Ti_{n+1}N_n$ have stronger preference to adhere to the terminal groups, which suggests more active surfaces for nitride-based MXenes. More importantly, nearly free electron states are observed to exist outside the surfaces of –OH functionalized carbide/nitride based MXenes, especially in $Ti_{n+1}N_n(OH)_2$, which provide almost perfect transmission channels without nuclear scattering for electron transport. The overall electrical conductivity of nitride-based MXenes is determined to be higher than that of carbide-based MXenes. The exceptional properties of titanium nitride-based MXenes, including strong surface adsorption, high elastic constant and Young's modulus, and good metallic conductivity, make them promising materials for catalysis and energy storage applications.

---

*Corresponding author; email: zaeem@mines.edu; zaeem@mst.edu (M. Asle Zaeem).

# 1. Introduction

One of the largest families of two-dimensional (2D) transition metal carbides/nitrides were recently introduced by selectively etching the A metal from MAX phases, $M_{n+1}AX_n$, where n is 1, 2 or 3, M is an early transition metal, A is a group IIIA or IVA element in the periodic table, and X is C and/or N [1, 2]. To emphasis their structural similarity to graphene, these 2D materials were named MXenes [1]. In recent years, 2D MXenes have attracted tremendous attention due to their great potential in a broad range of applications, including energy storage in Li-ion batteries (LIBs) [3-5] and hybrid electro-chemical capacitors (ECs) [4, 6], catalysis [7-9], hydrogen storage [10] and low-power electronics [11, 12].

Among all the experimentally synthesized and theoretically predicted MXenes, titanium carbide based MXenes are most widely studied, especially $Ti_3C_2T_2$ (T stands for the terminal groups: –F, –O and –OH), which was the first 2D MXene synthesized in 2011 [1]. In contrast, the synthesis of 2D titanium nitride-based MXenes is rather difficult due to the relative high formation energy of their three-dimensional (3D) MAX phases and their poor stability in the etchants, typically hydrofluoride (HF) [13]. After the first synthesis of $Ti_4N_3$ based MXene [14], a multilayered $Ti_2NT_x$ was also successfully obtained very recently by immersing $Ti_2AlN$ in a potassium fluoride (KF) – hydrochloric acid (HCl) mixture [15]. Although some progress have been made in synthesizing different 2D titanium carbide/nitride based MXenes, not enough systematic investigations have been completed to determine their structural, elastic and electronic properties, and accordingly to explore their potential applications.

The structure stability and surface chemistry of 2D materials are controlled by some basic structural features such as equilibrium lattice constants, monolayer thickness, and atomic composition. Because of the limitations of experimental approaches, the study of atomic scale structural features of monolayer MXenes are scarce; to the best of our knowledge, there are two experimental works that have been conducted to study the structural properties of $Ti_3C_2T_x$ monolayer flakes by scanning transmission electron microscope (STEM). The chemistry and surface kinetics [16], as well as intrinsic point defects [16, 17] of individual $Ti_3C_2T_x$ nanosheets at the atomic level were investigated. First principal calculations based on density functional theory (DFT) are alternative approaches to fundamentally study structure stability and surface chemistry of MXenes by calculating the cohesive energy and surface adsorption energy. Cohesive energy, which is defined as the energy required to separate the condensed material into isolated free atoms, is one of the most important physical parameters in quantifying the thermal stability of materials. However, most of the previous DFT studies were concentrated on examining the lattice parameter and layer thickness of titanium-based MXenes [18, 19], which are incapable of predicting the structural stability and surface properties of MXenes. There are only two DFT works [20, 21] that calculated the cohesive energies of MXenes, but they studied merely pristine MXenes; in other words, the effect of terminal groups on cohesive energy was ignored. Surface property of 2D materials are also very important for their applications in catalysis [7-9] and hydrogen storage [10]. For example, surface energy indicates how likely molecules are to adsorb (/desorb) onto the surface, or how strong the surface forms non-covalent bonds with other materials. Titanium, which is a transition metal, has been employed widely to increase the binding energy of hydrogen on carbon-based materials [22, 23], and the representative 2D $Ti_2C$ phase was proven to be an effective reversible hydrogen storage material [10]. Very recently, $Ti_3C_2$ MXene nanoparticle was explored and demonstrated to be a highly efficient co-catalyst [24]. MXenes were also investigated as electro catalysts for hydrogen evolution reaction (HER) [25]. However, up to now there are no computational reports on quantitatively elucidating the surface properties of 2D $Ti_{n+1}C_n$/$Ti_{n+1}N_n$ MXenes.

The elastic properties of 2D materials, such as the elastic constant ($C_{11}$), Young's modulus ($E$) and Poisson's ratio ($v$), are controlled by the high anisotropy between different crystallographic orientations of 2D materials, which is also responsible for their other unique optical, electronic and electrochemical properties [26]. Due to the challenges in experimental measurements of elastic properties of 2D MXenes at nanometer scale, a few computational studies, particularly by means of DFT calculations, have been carried out focusing on determining the elastic properties of some of the 2D MXenes. In the following, we briefly review the previous efforts for determining the elastic constants and Young's modulus of MXenes by DFT calculations and discuss some technical and/or physical issues associated with these works that have affected the reliability of their results. Kurtoglu et al. predicted the elastic constant ($C_{11}$) of some selected MXenes by applying a set of homogeneous deformations along the basal plane [27]. The calculated $C_{11}$ values of $Ti_2C$, $Ti_3C_2$ and $Ti_4C_3$ (636 GPa, 523 GPa and 512 GPa, respectively) were significantly smaller than that of graphene (1,028 GPa [28]), but they still demonstrated that 2D $Ti_{n+1}C_n$ solids are quite stiff. A technical issue associated with this work was that the interlayer spacing to create a 2D layer of atoms was chosen to be half of the distance that was used in other DFT studies of 2D MXenes (other works used 20 Å interlayer spacing) [10, 19, 29, 30]. A small spacing in-between layers leads to fluctuation in energy and ultimately affect the results, and to eliminate the interaction between free surfaces in DFT, a larger spacing is needed. A physical issue associated with this work was that the effect of Poisson's ratio was neglected, and instead the lattice vector in the transverse direction was controlled at a constant value by adopting periodic boundary conditions. Such treatment is akin to applying a constraint force on the transverse section during uniaxial tension or compression, which will consequently result in overestimating the elastic constants. In a different work but using the method of calculation in [27], the stress-strain relations for 2D $Ti_2C$ under both biaxial and uniaxial load conditions were obtained by Guo et al. [29], showing that 2D $Ti_2C$ is elastically isotropic, and the calculated Young's modulus along two perpendicular directions (zigzag and armchair directions) were 620 GPa and 600 GPa, respectively. By using the same loading and boundary conditions but different interlayer spacing, Chakraborty et al. [30] recently obtained a similar Young's modulus (~580 GPa) to Guo et al. work [29], but a higher elastic constant (~710 GPa) than Kurtoglu et al. work [27] for 2D $Ti_2C$. As aforementioned, such unrelaxed transverse boundary condition can lead to overestimation of elastic properties. Moreover, to the best of our knowledge, the elastic properties of pristine $Ti_{n+1}N_n$ have not yet studied. As part of this work, we revisit DFT calculations of the elastic properties of pristine $Ti_{n+1}C_n$ MXenes in order to improve the reliability of the results, and also investigate the elastic properties of pristine $Ti_{n+1}N_n$.

In addition to structural and mechanical properties, knowledge of electronic properties is also greatly desired to determine metallic conductivity of MXenes. To assure efficient charge-carrier transfer during energy storage, excellent metallic conductivity of these MXenes is highly desired. During the etching process of MAX phases, the MXene surfaces acquire numerous hydrophilic terminal groups, which render them as promising catalyst materials. However, the presence of surface functional groups may lead to dramatic changes in the electronic properties of MXenes. For example, previous first principles calculations on $M_2C$ based MXenes have shown that upon appropriate surface functionalization, some of the MXenes such as $Ti_2CO_2$, $Hf_2CO_2$, $Zr_2CO_2$ and $Sc_2CO_2$ become semiconducting, which is differ from their bare transition metal carbide monolayers that show high metallic conductivity [31, 32]. However, little work has been done on the electronic properties of 2D $Ti_{n+1}N_nT_2$, $Ti_3C_2T_2$ and $Ti_4C_3T_2$ MXenes. Interesting open questions are arising that how the terminal groups interact with the surface atoms of pristine $Ti_{n+1}C_n$ and $Ti_{n+1}N_n$, how they affect the metallic conductivity of different MXenes, and do $Ti_{n+1}N_nT_2$ possess as good as or better metallic conductivity than $Ti_{n+1}C_nT_2$? To answer these questions, as part of this work, we study the electronic properties of $Ti_{n+1}C_nT_2$ and $Ti_{n+1}N_nT_2$ (n=1, 2, 3) by DFT calculations.

In the present study, we extensively investigate the structural, elastic and electronic properties of 2D $Ti_{n+1}C_n$ and $Ti_{n+1}N_n$ (n=1, 2 and 3) based MXenes by performing first principles calculations. Lattice parameter, layer thickness, cohesive energy, adsorption energy of terminal groups onto the surface of MXenes, elastic constant, Young's modulus, Poisson's ratio, and electrical conductivity are determined for bulk TiC/TiN, 2D pristine $Ti_{n+1}C_n$/$Ti_{n+1}N_n$ (n=1, 2 and 3), and functionalized $Ti_{n+1}C_nT_2$/$Ti_{n+1}N_nT_2$ (T = –F, –O and –OH) MXenes. The effect of terminal groups on the chemical bonding between elements in 2D carbide/nitride based MXenes are also analyzed by electron localization function (ELF) and charge density distribution (CDD). Finally, their density of states (DOS) are calculated to characterize their metallic conductivities.

## 2. Computational Models and Calculation Details

First principles calculations are performed using DFT in conjunction with projector augmented wave (PAW), as implemented in the Vienna Ab inito Simulation Package (VASP) [33]. The exchange-correlation energy of interacting electrons is treated by both the Perdew-Burke-Ernzerhof (PBE) version of the generalized gradient approximation (GGA) [34]. A plane wave cutoff energy of 400 eV is found to be sufficient to ensure that the total energies are converged to less than $10^{-4}$ eV/unit cell. The *k*-point of 9×5×1 is used for structure optimization and static self-consistent calculation for $Ti_{n+1}X_n$ (X = C, N) monolayers.

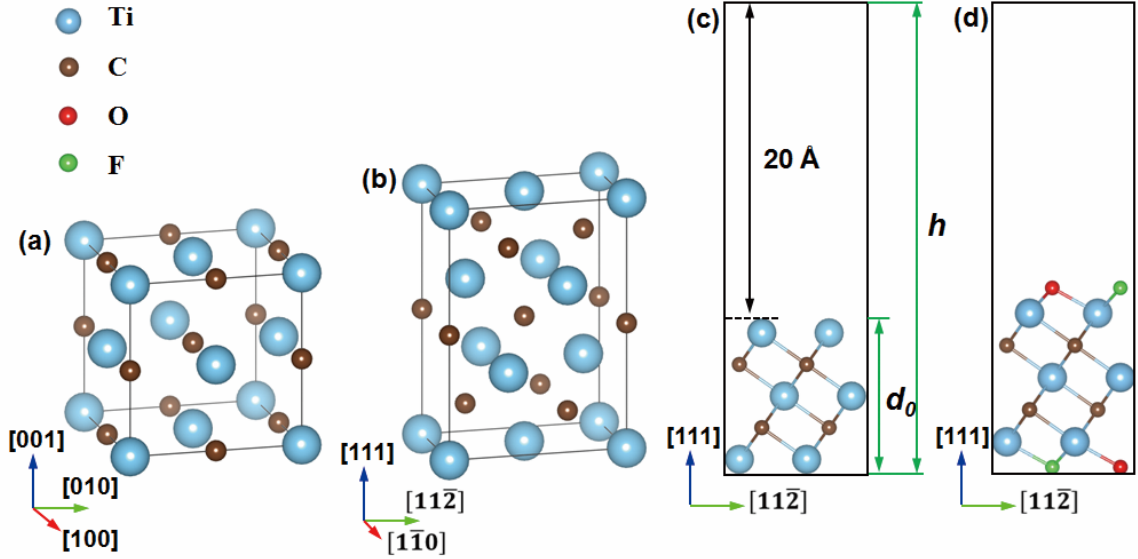

Figure 1. (a) Crystal structure of bulk TiC; (b) Periodic boundary condition in [111] direction is applied on layered $Ti_{n+1}C_n$ to model the corresponding bulk material. (c) Side view of a slab model for pristine $Ti_3C_2$ with a vacuum space of 20 Å. (d) Side view of $Ti_3C_2$ slab model with hybrid surface terminations: –O and –F.

Figure 1a present the crystal structure of bulk TiC. The lateral plane of 2D $Ti_{n+1}C_n$ is parallel to the (111) surface of bulk TiC. In Figure 1b, by applying periodic boundary conditions, the bulk state of $Ti_{n+1}C_n$ can be obtained, which is denoted as bulk $Ti_{n+1}C_n$ - $[1\bar{1}0]$ to differentiate from the bulk TiC in Figure 1a. To simulate MXene sheets, we build slab models, in which a large vacuum space of 20 Å is added above the upper layer to prevent its interaction with the periodic image of the lower layer, as shown in Figure 1c. In our studies, we have also considered the cases of functionalized MXenes, i.e., surfaces of $Ti_{n+1}X_n$ monolayers are fully terminated by –F, –O and –OH. Figure 1d shows an example of $Ti_3C_2$ with hybrid –O and –F terminations.

Lattice constant and layer thickness are important parameters in characterizing the crystal configurations of MXenes. In terms of structural stability, the cohesive energy $E_{coh}$, being a measure of the strength of forces that bind atoms together in a crystal, is a useful parameter in understanding the phase stability [20, 21]. In general, the cohesive energy is defined as the total energy of the compound deducted by the sum of the energies of all the individual constituent atoms [20]. In the case of $Ti_{n+1}X_nT_2$ MXenes, the cohesive energy is calculated by the following formula (Equation (1)):

$$E_{coh} = E_{tot}(Ti_{n+1}X_nT_2) - (n+1)E_{atm}(Ti) - nE_{atm}(X) - 2E_{atm}(T), \tag{1}$$

where $E_{tot}$ is the total energy of $Ti_{n+1}X_nT_2$, $E_{atm}$ is the energy of free atoms of Ti, X (C/N), and T (–O, –F or –OH). To normalize the cohesive energy of different systems, we calculate the cohesive energy per atom by: $\bar{E}_{coh} = E_{coh}/(n+1+n+2)$ (for pristine $Ti_{n+1}X_n$, $\bar{E}_{coh} = E_{coh}/(n+1+n)$).

In addition, the stability trend of competing phases, such as $Ti_2C$, $Ti_3C_2$, and $Ti_4C_3$, can be obtained by calculating their formation energies. This approach has successfully analyzed the stability trends of some MAX phases previously [35, 36]. In the current study, for pristine $Ti_{n+1}C_n$ and $Ti_{n+1}N_n$, the reactions are assumed to be $Ti_{n+1}C_n \leftrightarrow Ti_nC_{n-1} + TiC$, and $Ti_{n+1}N_n \leftrightarrow Ti_nN_{n-1} + TiN$, where TiC and TiN have B1 crystal structures. Therefore, the formation energy for $Ti_{n+1}C_n$ can be calculated by: $\Delta E_{form}(Ti_{n+1}C_n) = E_{tot}(Ti_{n+1}C_n) - E_{tot}(Ti_nC_{n-1}) - E_{tot}(TiC)$, and the formation energy for $Ti_{n+1}N_n$ can be calculated in a similar way. Within this definition, a negative $\Delta E_{form}$ indicates that it is energetically favorable for $Ti_nC_{n-1}$ to form the stable phase of $Ti_{n+1}C_n$, and *vice versa*.

In general, surface energy can be determined by taking the energy difference between the total energy of a slab model and an equivalent bulk reference. However, for MXenes since the slab model are terminated with two identical surfaces (Ti atoms), the task of calculating the surface energy is challenging, because there is no appropriate approach for building a bulk material by repeating the MXene monolayer along its thickness direction. In order to quantify the chemical bonding strength between Ti atoms at the surface and terminal groups (–F, –O and –OH), we further calculate the adsorption energy ($E_{ads}$) of the terminal groups on the surfaces of $Ti_{n+1}C_n$ and $Ti_{n+1}N_n$ monolayers. The adsorption energy is defined as:

$$E_{ads} = \frac{E_{tot(Ti_{n+1}X_nT_2)} - (E_{Ti_{n+1}X_n} + E_{T_2})}{2}, \quad (2)$$

where $E_{tot(Ti_{n+1}X_nT_2)}$ stands for the total energy of the functionalized MXenes, $E_{Ti_{n+1}X_n}$ and $E_{T_2}$ are the total free energy of the pristine $Ti_{n+1}X_n$ monolayer and the isolated terminal group atoms, respectively. *2* in the denominator of Equation (2) indicates the number of adsorbed terminal groups on the surfaces, which means there are top and bottom Ti surfaces.

In VASP by updating the symmetry inequivalent displacement, the elastic constants can be calculated automatically. Then, the elastic tensor is determined by performing six finite distortions of the lattice and deriving the elastic constants from the stress-strain relationship. Since the force is averaged over the entire simulation cell, including the vacuum space, the elastic constant is rescaled by $h/d_0$ to obtain the effective elastic constants, where $h$ is the slab model length in the [111] direction and $d_0$ is the thickness of the monolayer. The elastic mechanical response is determined by an equivalent continuum mechanics approach [28, 37], in which the stiffness tensor (elastic constants), Young's modulus, Poisson's ratio and shear modulus are inversely related as shown in Equation (3):

$$\begin{bmatrix} C_{11} & C_{12} & C_{13} & 0 & 0 & 0 \\ & C_{22} & C_{23} & 0 & 0 & 0 \\ & & C_{33} & 0 & 0 & 0 \\ & & & C_{44} & 0 & 0 \\ & Symm & & & C_{55} & 0 \\ & & & & & C_{66} \end{bmatrix}^{-1} = \begin{bmatrix} \frac{1}{E_1} & \frac{-v_{12}}{E_1} & \frac{-v_{13}}{E_1} & 0 & 0 & 0 \\ & \frac{1}{E_2} & \frac{-v_{23}}{E_2} & 0 & 0 & 0 \\ & & \frac{1}{E_3} & 0 & 0 & 0 \\ & & & \frac{1}{G_{23}} & 0 & 0 \\ & Symm & & & \frac{1}{G_{13}} & 0 \\ & & & & & \frac{1}{G_{12}} \end{bmatrix}, \quad (3)$$

where $E_i$ is the Young's modulus in the $i$ direction, $v_{ij}$ is the Poisson's ratio associated with the $i$ and $j$ directions, and $G_{ij}$ is the shear modulus in the $ij$ plane. For the case of 2D monolayers, the most concern of mechanical response is focused on the lateral plane. Moreover, our current study as well as the previous literatures [29, 30], demonstrate that $Ti_{n+1}C_n$ MXenes are nearly elastically isotropic. Therefore, the in-plane elastic constant ($C_{11}$), Young's modulus ($E$) and Poisson's ratio ($v$) are investigated in the current study. Considering the reduced dimensionality, the in-plane Young's modulus ($E$) and Poisson's ratio ($v$) can be obtained from the following relationships: $E = (C_{11}^2 - C_{12}^2)/C_{11}$ and $v = C_{12}/C_{11}$ [38, 39].

ELF is a very powerful tool in categorizing and evaluating the chemical bonding between elements in the material. To gain a direct observation of the chemical bonding change of $Ti_{n+1}X_n$ based MXenes (X = C, N), we further analysis ELF and electron charge density distributions (CDD). VESTA [40] is used to visualize the contours of ELF and CDD. ELF is defined as [41]:

$$ELF = 1/[1+(D/D_h)] \quad (4)$$

$$D = \frac{1}{2}\sum_i |\nabla \varphi_i|^2 - \frac{1}{8}\frac{|\nabla \rho|^2}{\rho} \tag{5}$$

$$D_h = \frac{3}{10}(3\pi^2 \rho)^{\frac{5}{3}} \tag{6}$$

$$\rho = \sum_i |\varphi_i|^2 \tag{7}$$

where $\varphi_i$ represents the Kohn-Sham orbitals, and $\rho$ stands for the electron charge density.

Total and partial DOS of pristine $Ti_{n+1}C_n/Ti_{n+1}N_n$ and functionalized $Ti_{n+1}C_nT_2/Ti_{n+1}N_nT_2$ are also analyzed to better understand their electrical conductivity.

## 3. Results and Discussion

### *3.1 Structural properties*

In order to investigate the ground-state properties, we simulated the equilibrium configurations of pristine $Ti_{n+1}C_n/Ti_{n+1}N_n$ and their functionalized monolayers. Since to date there are no available experimental data for monolayer MXenes to compare to DFT calculations at nano- or electronic scale, we also performed DFT calculations of bulk TiC and bulk TiN and compared the results to experimental results in literatures to validate the accuracy of our method. The calculated lattice constants and monolayer thicknesses are listed in Table 1, in comparison with available experimental and theoretical data reported in the literatures. Firstly, it can be seen that our calculated lattice constants of bulk TiC and TiN are in very good agreement with the results from other experimental works [42-44]. Secondly, $Ti_3C_2$ and $Ti_3N_2$ possess a relatively larger lattice constant comparing to $Ti_2C$, $Ti_4C_3$, $Ti_2N$, and $Ti_4N_3$. Additionally, we find that the lattice constants and monolayer thicknesses of carbide-based MXenes are larger than those of nitride-based MXenes, which is correlated with the atomic radii difference between carbon and nitrogen.

Table 1. Calculated lattice constant $a$ (Å) and layer thickness $d$ (Å) of 2D pristine $Ti_{n+1}C_n$/$Ti_{n+1}N_n$, functionalized $Ti_{n+1}C_nT_2$/$Ti_{n+1}N_nT_2$ (T = –F, –O, –OH; n = 1, 2, 3), and bulk TiC/TiN, in comparison with available experimental and first-principles data from the literature (inside the parentheses).

| $Ti_{n+1}C_nT_2$ | $a$ (Å) | $d$ (Å) | $Ti_{n+1}N_nT_2$ | $a$ (Å) | $d$ (Å) |
|---|---|---|---|---|---|
| Bulk TiC | 4.32(4.33,[42] 4.32,[43] 4.34[44]) | – | Bulk TiN | 4.24 (4.24,[42] 4.26[44]) | – |
| $Ti_2C$ | 3.04 (3.01,[27] 3.04,[19] 3.04[20]) | 2.31 (2.29,[27] 2.31[20]) | $Ti_2N$ | 2.98 (2.98,[21] 2.9853[20]) | 2.29 (2.29[20]) |
| $Ti_2CF_2$ | 3.04 | 4.80 | $Ti_2NF_2$ | 2.98 | 4.76 |
| $Ti_2CO_2$ | 3.04 | 4.45 | $Ti_2NO_2$ | 2.98 | 4.47 |
| $Ti_2C(OH)_2$ | 3.04 | 6.83 | $Ti_2N(OH)_2$ | 2.98 | 6.78 |
| $Ti_3C_2$ | 3.10 (3.07,[27] 3.09[20]) | 4.64 (4.61,[27] 4.74[20]) | $Ti_3N_2$ | 3.05 (3.01[20]) | 4.73 (4.56[20]) |
| $Ti_3C_2F_2$ | 3.10 | 7.18 | $Ti_3N_2F_2$ | 3.05 | 7.18 |
| $Ti_3C_2O_2$ | 3.10 | 6.87 | $Ti_3N_2O_2$ | 3.05 | 6.80 |
| $Ti_3C_2(OH)_2$ | 3.10 | 9.20 | $Ti_3N_2(OH)_2$ | 3.05 | 9.22 |
| $Ti_4C_3$ | 3.09 (3.07,[27] 3.10[20]) | 7.15 (7.14,[27] 7.15[20]) | $Ti_4N_3$ | 2.99 (2.99[14, 20]) | 7.36 (7.22,[20] 7.44[14]) |
| Ti$_4$C$_3$F$_2$ | 3.09 | 9.65 | Ti$_4$N$_3$F$_2$ | 2.99 | 9.74 |
| Ti$_4$C$_3$O$_2$ | 3.09 | 9.36 | Ti$_4$N$_3$O$_2$ | 2.99 | 9.38 |
| Ti$_4$C$_3$(OH)$_2$ | 3.09 | 11.67 | Ti$_4$N$_3$(OH)$_2$ | 2.99 | 11.74 |

The calculated normalized cohesive energies for pristine Ti$_{n+1}$C$_n$/Ti$_{n+1}$N$_n$ and functionalized Ti$_{n+1}$C$_n$T$_2$/Ti$_{n+1}$N$_n$T$_2$ are listed in Table S1 and compared in Figure 2. It can be seen that our results for pristine Ti$_{n+1}$C$_n$/Ti$_{n+1}$N$_n$ are in good agreement with previous studies [20, 21], which certifies the reliability of our present DFT calculations. The obtained $\bar{E}_{coh}$ for all the studied MXenes are relatively small negative values, i.e., around $-6.0 \sim -7.0$ eV/atom, as shown in Table S1, which demonstrates that all the considered MXenes have stable structures. Moreover, the values of $\bar{E}_{coh}$ for pristine Ti$_{n+1}$C$_n$/Ti$_{n+1}$N$_n$ decrease with increasing $n$, i.e., increasing the thickness of monolayers (Figure 2). In other words, structural stability increases in these sequences: Ti$_2$C < Ti$_3$C$_2$ < Ti$_4$C$_3$ and Ti$_2$N < Ti$_3$N$_2$ < Ti$_4$N$_3$, and this is due to the increase in number of stronger Ti-C/N bonds in thicker MXene monolayers. Our calculated formation energies are $\Delta E_{form}(\text{Ti}_3\text{C}_2) = -0.283$ eV, $\Delta E_{form}(\text{Ti}_4\text{C}_3) = -0.005$ eV, $\Delta E_{form}(\text{Ti}_3\text{N}_2) = -0.187$ eV, and $\Delta E_{form}(\text{Ti}_4\text{N}_3) = -0.003$ eV. These negative values of $\Delta E_{form}$ confirm the obtained stability trends of Ti$_2$C < Ti$_3$C$_2$ < Ti$_4$C$_3$ and Ti$_2$N < Ti$_3$N$_2$ < Ti$_4$N$_3$, which agrees with the conclusions made from their cohesive energy calculations. It is also worth to mention that $\Delta E_{form}$ of Ti$_{n+1}$N$_n$ are smaller than those of Ti$_{n+1}$C$_n$, thus the preparation of free-standing Ti$_{n+1}$N$_n$ will be more problematic, which is consistent with the experimental observations [14, 15].

Similar stability trends are also observed in functionalized MXenes. In terms of the effect of functionalization, –O terminated MXenes possess the lowest $\bar{E}_{coh}$ compared to others terminated with –F and –OH, from which we can conclude that structural stability of functionalized MXenes decreases in these sequences: $Ti_{n+1}C_nO_2$ > $Ti_{n+1}C_nF_2$ > $Ti_{n+1}C_n(OH)_2$ and $Ti_{n+1}N_nO_2$ > $Ti_{n+1}N_nF_2$ > $Ti_{n+1}N_n(OH)_2$. Such conclusion is consistent with the nuclear magnetic resonance (NMR) spectroscopy results [45] that the amount of –OH terminations is significantly fewer than –F and –O in $Ti_3C_2T_x$ MXenes. In general, the value of $\bar{E}_{coh}$ is significantly decreased through surface functionalization, particularly by –O and –F groups, which indicates that the fully functionalized MXenes are more stable than their corresponding pristine phases. This also provides an alternative theoretical explanation of why the experimental synthesized MXenes are usually functionalized with terminal groups [2]. Additionally, it is noted that the overall values of $\bar{E}_{coh}$ of $Ti_{n+1}C_n$ and $Ti_{n+1}C_nT_2$ are lower than those of $Ti_{n+1}N_n$ and $Ti_{n+1}N_nT_2$ with same monolayer thickness (same value of *n*), which indicates that carbide-based MXenes are more stable than nitride-based ones. This conclusion is in good agreement with the experimental observations which have shown that nitride-based MXenes have poor stability in the etchant during the synthesis process [13, 15].

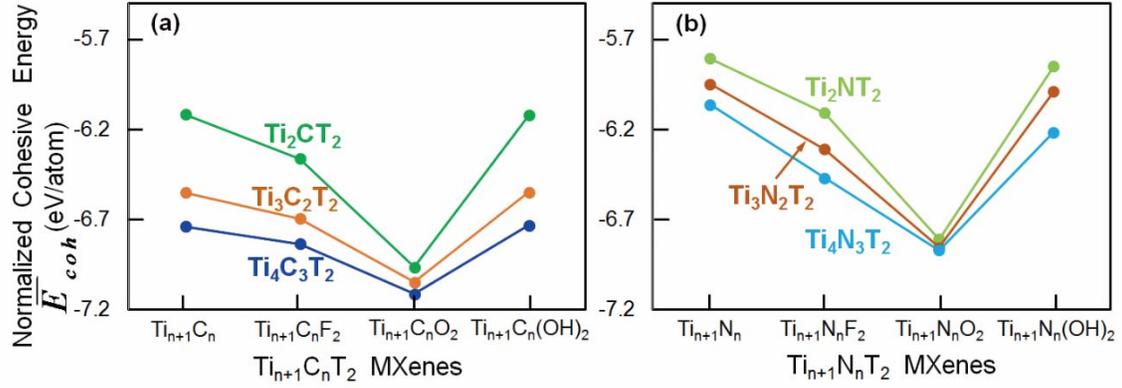

Figure 2. Calculated normalized cohesive energies $\bar{E}_{coh}$ (eV/atom) of pristine $Ti_{n+1}C_n/Ti_{n+1}N_n$ and functionalized $Ti_{n+1}C_nT_2/Ti_{n+1}N_nT_2$ (T = –F, –O, –OH; n = 1, 2, 3) MXene monolayers.

The calculated adsorption energies for adhesion of terminal groups onto the surface of pristine $Ti_{n+1}C_n$ and $Ti_{n+1}N_n$ monolayers are listed in Table S2 and plotted in Figure 3 for comparison. It is noticed that the overall adsorption energies of $Ti_{n+1}N_n$ are lower than those of $Ti_{n+1}C_n$, which implies that $Ti_{n+1}N_n$ has more active surface chemistry than $Ti_{n+1}C_n$. It has been reported that the difficulty in producing titanium nitride-based MXenes is caused partially by the strong metallic bonding between Ti-A atoms in their MAX phase (for example Ti-Al in $Ti_{n+1}AlN_n$), and thus requiring more energy to extract the A atoms [13, 14]. Analogously, our adsorption energy calculation demonstrate that terminal groups prefer to adhere to the surfaces of $Ti_{n+1}N_n$ rather than $Ti_{n+1}C_n$. Besides, –O terminated MXenes ($Ti_{n+1}C_nO_2$ and $Ti_{n+1}N_nO_2$) are found to possess the minimum values of adsorption energies, while –OH terminated MXenes ($Ti_{n+1}C_n(OH)_2$ and $Ti_{n+1}N_n(OH)_2$) show the maximum values. Therefore, we can conclude that –O groups have the most preference to be adhered onto the surfaces of the pristine $Ti_{n+1}C_n$ and $Ti_{n+1}N_n$ during the etching process. This conclusion is in good agreement with the afore analyses of $\bar{E}_{coh}$. The obtained trend of adsorption energies of $Ti_{n+1}C_n$ based MXenes is in good agreement with a previous DFT study [46]. Moreover, our results are consistent with experimental results regarding the stoichiometry of terminations in $Ti_3C_2T_x$ obtained by mild etchant [45].

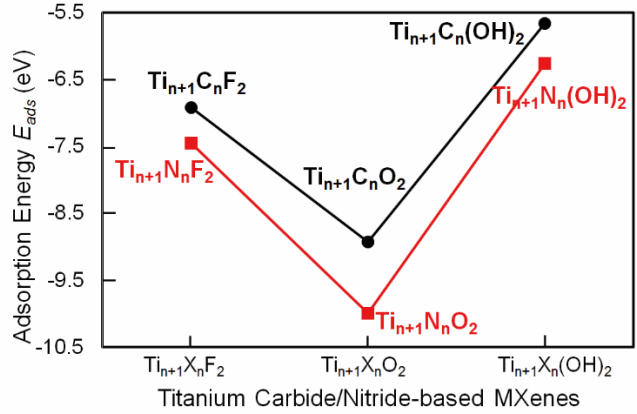

Figure 3. The adsorption energies ($E_{ads}$) for adhesion of terminal groups (-F, -O, and -OH) onto the surfaces of $Ti_{n+1}C_n$ and $Ti_{n+1}N_n$.

*3.2 Elastic properties*

The calculated elastic properties of pristine $Ti_{n+1}C_n$ and $Ti_{n+1}N_n$ and bulk TiC and TiN are presented in Table 2. It should be noted that since 2D titanium carbide is elastically isotropic [29], in the current study only elastic properties along armchair direction of pristine $Ti_{n+1}C_n$ and $Ti_{n+1}N_n$ are calculated. The results for bulk TiC and TiN are much closer to the experimental measurements in comparison with the other first-principles calculations. Therefore, it can be concluded that the present DFT method is able to provide reliable results for pristine $Ti_{n+1}C_n$ and $Ti_{n+1}N_n$. The results indicate that the calculated Young's moduli of $Ti_{n+1}C_n$ decrease with the increase of monolayer thickness, i.e., in the order of $Ti_2C > Ti_3C_2 > Ti_4C_3$ with values of 601 GPa, 473 GPa and 459 GPa, respectively; a similar trend was also reported by previous DFT calculations for Ti$_{n+1}$C$_n$ (n = 1, 2, 3) [29]. Therefore, in general one may infer that Young's modulus of MXenes decreases with the increase of the monolayer thickness. To confirm this conclusion, we increase the layer of atoms to infinite to generate a bulk model, as shown in Figure 1b. The calculated Young's modulus for bulk Ti$_{n+1}$C$_n$ is 433 GPa, which is smaller than all three Ti$_{n+1}$C$_n$. Similar trend is also observed in Ti$_{n+1}$N$_n$, however, the Young's modulus of Ti$_{n+1}$N$_n$ is higher than that of Ti$_{n+1}$C$_n$, as shown in Table 2. These results are reasonable because the experimental measurements show that the Young's modulus of bulk TiN [47] is larger than that of bulk TiC [48, 49]. The in-plane Poisson's ratio ($v$) is also calculated. It is noted that due to the ultra-thin thickness of Ti$_{n+1}$C$_n$ and Ti$_{n+1}$N$_n$, their Poisson's ratios are around 0.25 and 0.26, which are larger than that of the bulk TiC and TiN (~0.23) [47-49].

Table 2. Summary of the in-plane elastic properties of bulk TiC/TiN and 2D pristine Ti$_{n+1}$C$_n$/Ti$_{n+1}$N$_n$ (n = 1, 2, 3) monolayers, including elastic constant $C_{11}$ (GPa), Young's modulus $E_1$ (GPa), and Poisson's ratio $v$. For comparison, the available experimental and first-principles data are also listed inside the parentheses.

|  | $C_{11}$ (GPa) | $E_1$ (GPa) | $v$ |
| --- | --- | --- | --- |
| Bulk TiC | 577 (513,[50] 606[51]) | 524 (450[48, 49]) | 0.237 |
| Bulk Ti$_{n+1}$C$_n$ | 433 | 379 | 0.285 |
| Ti$_2$C | 601 (636[27]) | 513 (620, 600[29]) | 0.266 |
| Ti$_3$C$_2$ | 473 (523[27]) | 447 | 0.241 |
| Ti$_4$C$_3$ | 459 (512[27]) | 431 | 0.250 |
| Bulk TiN | 647 (625,[52] 713[53]) | 594 (490[47]) | 0.231 |
| Bulk Ti$_{n+1}$N$_n$ | 562 | 456 | 0.312 |
| Ti$_2$N | 654 | 538 | 0.271 |
| Ti$_3$N$_2$ | 557 | 504 | 0.265 |
| Ti$_4$N$_3$ | 501 | 471 | 0.250 |

*3.3 Electronic properties*

To gain a direct observation of the change of chemical bonding, we first analyze the ELF of pristine Ti$_{n+1}$C$_n$/Ti$_{n+1}$N$_n$ and functionalized Ti$_{n+1}$C$_n$T$_2$/Ti$_{n+1}$N$_n$T$_2$. Figure 4a-d and Figure 4a'-d' display the effect of functionalization on the ELF contour plots (projected in the ($1\bar{1}0$) plane) of Ti$_{n+1}$C$_n$ and Ti$_{n+1}$N$_n$ based MXenes, respectively. The regions close to the unity (red areas) contain many localized electrons, which indicates a region around a nucleus or in a very strong covalent bonding condition. Values close to zero (blue areas) represent the regions with low electron density, and the values close to 0.5 (green areas) correspond to a uniform electron gas where the bonding might have a metallic character.

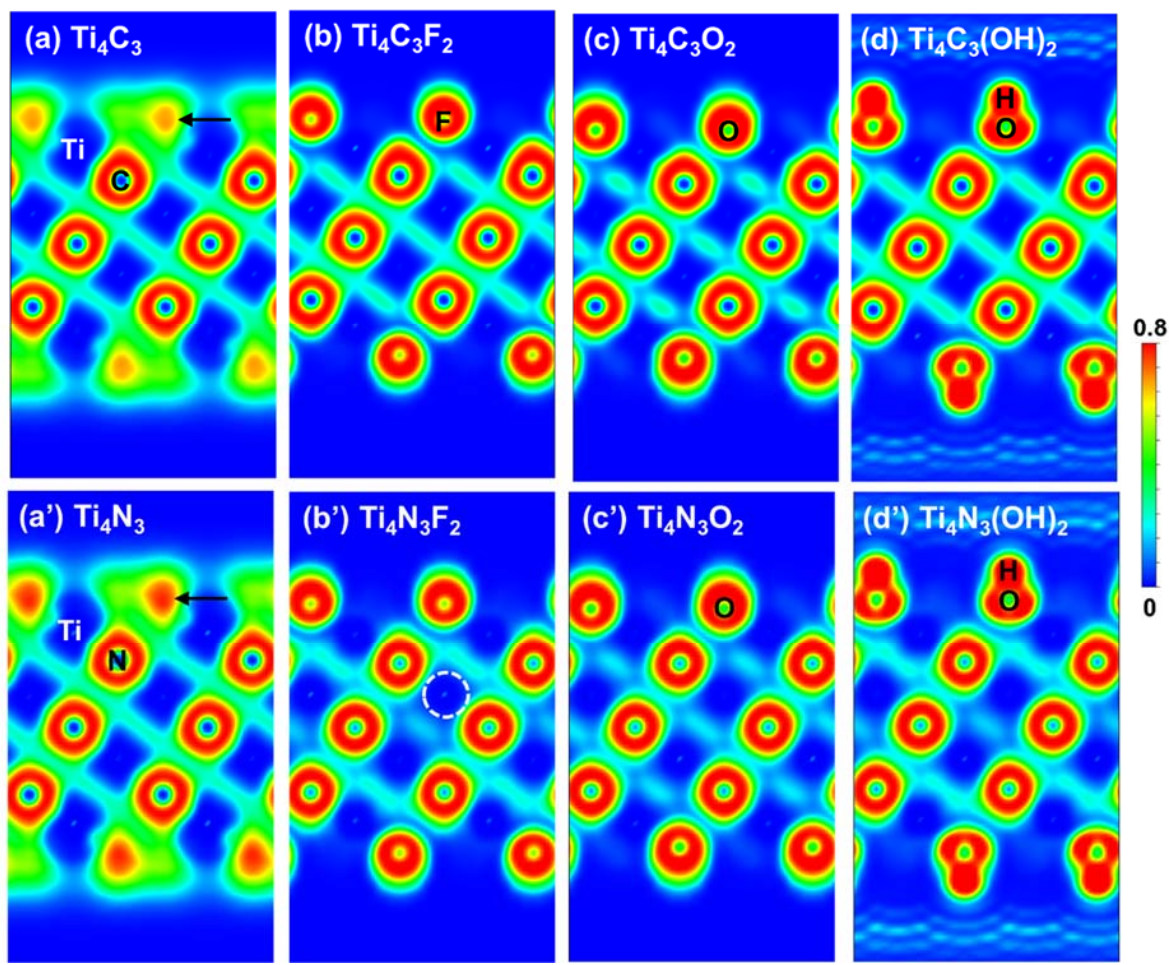

Figure 4. The ELF contour plots for (a-d) $Ti_4C_3$, $Ti_4C_3F_2$, $Ti_4C_3O_2$, and $Ti_4C_3(OH)_2$, and (a'-d') $Ti_4N_3$, $Ti_4N_3F_2$, $Ti_4N_3O_2$, and $Ti_4N_3(OH)_2$ MXenes.

It can be noted from Figure 4 that a substantial concentration of electrons is located around C/N atoms and the terminal groups –F, –O and –OH. Besides, for pristine $Ti_4C_3$ and $Ti_4N_3$, high density localized electrons with ELF values of 0.65~0.8, as denoted by black arrows in Figure 4, are observed, which are composed of lone pair electrons donated by the surface Ti atoms. Such delocalized nature of electrons are also observed in $Cr_2C$ MXenes [54] and on the surfaces of the other pristine titanium carbide (Figure S1a, a' and a'') and nitride MXenes (Figure S2a, a' and a''). However, the density of lone pair electrons in $Ti_4N_3$ (Figure 4a) is higher than that in $Ti_4C_3$ (Figure 4a'). With the addition of terminal groups those lone pair electrons disappear. Instead, dense electron cloud is observed to surround both C/N atoms and the terminal groups, except that the density of dispersing electron gas between N-N atoms (light blue color in Figure 4) is relatively lower than that between C-C atoms, which implies a weaker covalent bonding. Especially, it is interesting to note that different terminal groups induce various electronic structures. –F and –OH terminated MXenes show very similar behaviors but different from the ones terminated with –O group. This is because each of the –F and –OH groups can receive only one electron from each surface Ti atom, while –O group demands two electrons in order to be stabilized at its location of adsorption. Comparing to $Ti_4C_3T_2$, the electrons around Ti atoms in $Ti_4N_3T_2$ become more localized, which leads to the formation of local spherical geometries, as denoted in 2D representation in Figure 4 by a dashed circle. Such variations indicate that the ionic bonding interactions between Ti-N become stronger due to functionalization.

Moreover, ELF analyses also reveal the existence of nearly free electron (NFE) states in –OH terminated MXenes, e.g., $Ti_4C_3(OH)_2$ and $Ti_4N_3(OH)_2$ (Figure 4d, d'). In general, these NFE states are spatially located outside the surface atoms and are extended parallel to the surfaces. We have also calculated the projected band structures of –OH terminated MXenes to confirm the existence of NFE states, as shown in Figure 5 and Figure S3. Clearly, the band structures show NFE characteristics with parabolic energy dispersions with respect to the crystal wave vector, as indicated by red arrows in Figure 5 and Figure S3, which is consistent with the observations in ref. [41]. It has been shown that the NFE states concentrate in the region of highest positive charges [55]. As a results, we found uniform floating electron gas with ELF $\cong$ 0.3 exist above the hydrogen atoms in $Ti_{n+1}C_n(OH)_2$ and $Ti_{n+1}N_n(OH)_2$, which is attributed to the positive charge of H atoms. Although the distribution of these NFE states is insensitive to the monolayer thickness (see Figure S1d, d', d'' and Figure S2d, d', d''), their density is noted to increase with the increase of monolayer thickness in $Ti_{n+1}C_n(OH)_2$. Electron transport calculations have demonstrated that the NFE states provide almost perfect transmission channels without nuclear scattering for electron transport [41]. Moreover, the existence of such trapped states on the surface might promote the catalytic properties [56]. Owing to the stronger bonding of Ti-N than Ti-C, the overall density of the NFE states in $Ti_{n+1}N_n(OH)_2$ (Figure 4d' and Figure S2) is higher than that of $Ti_{n+1}C_n(OH)_2$ (Figure 4d and Figure S1). Hence, $Ti_{n+1}N_n(OH)_2$ MXenes have much more potential applications in nanoelectronic devices, catalysts and sensors than $Ti_{n+1}C_n(OH)_2$ MXenes.

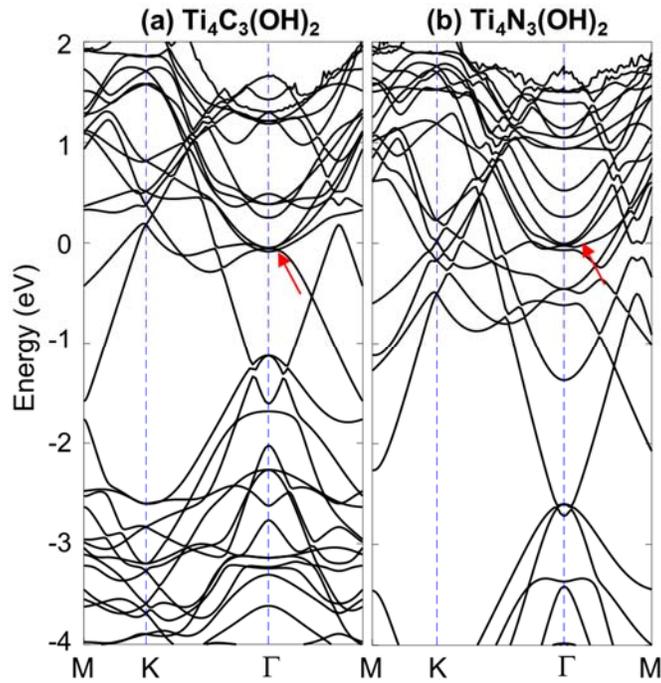

Figure 5. Projected energy band structures of –OH terminated MXenes: (a) $Ti_4C_3(OH)_2$ and (b) $Ti_4N_3(OH)_2$. The red arrows indicate the lowest energy NFE near the Fermi energy at the Γ point.

To directly visualize the bonding interaction between atoms, we also plot CDD of pristine $Ti_{n+1}C_n/Ti_{n+1}N_n$ and functionalized $Ti_{n+1}C_nT_2/Ti_{n+1}N_nT_2$ in Figure 6. It can be noted that although there are slight electron cloud overlap between the surface Ti and C/N atoms in pristine $Ti_4C_3/Ti_4N_3$ (highlighted by arrows in Figure 6a, a'), ionic bonding is the essential interaction. Monolayer thickness shows a negative effect on such ionic bonding in pristine $Ti_{n+1}C_n/Ti_{n+1}N_n$, in other words, partial ionic bonding transfers to covalent bonding with the increase of monolayer thickness (see Figure S4 and Figure S5). After functionalization, ultra-high CDD is observed around the nuclei of terminal groups, as well as N atoms (Figure 6b'-d'). N nuclei are surrounded by the high CDD ($\approx 0.25$), and the thickness of electron gas (CDD $\approx 0.15$) is reduced due to the addition of terminal groups, which is different from the C nuclei in $Ti_4C_3T_2$. The electron clouds outside the Ti nuclei, particularly in $Ti_4C_3O_2$, are significantly stretched along the Ti-O and Ti-C bond paths, which results in an approximate diamond geometry, as denoted in Figure 6c. In the case of $Ti_4N_3O_2$ (Figure 6c'), CDD of surface Ti is slightly elongated, and weak covalent bonding forms between outer layer Ti-O, similar to the case of $Ti_4C_3O_2$. Such unique behavior can be used to explain ultra-low adsorption energy of –O group onto the surfaces of $Ti_{n+1}C_n/Ti_{n+1}N_n$. However, the inner layer Ti atoms of $Ti_4N_3O_2$, as well as those in $Ti_4N_3F_2$ and $Ti_4N_3(OH)_2$ possess circular shape CDDs, and no sharing of electron cloud is observed, as denoted by dashed circles in Figure 6b'-d'. Such phenomenon implies that the original covalent boding between Ti-N partially or completely transfer to an essential ionic interaction due to the functionalization by –O, –F and –OH groups, which is also consistent with the conclusion deduced from ELF analysis.

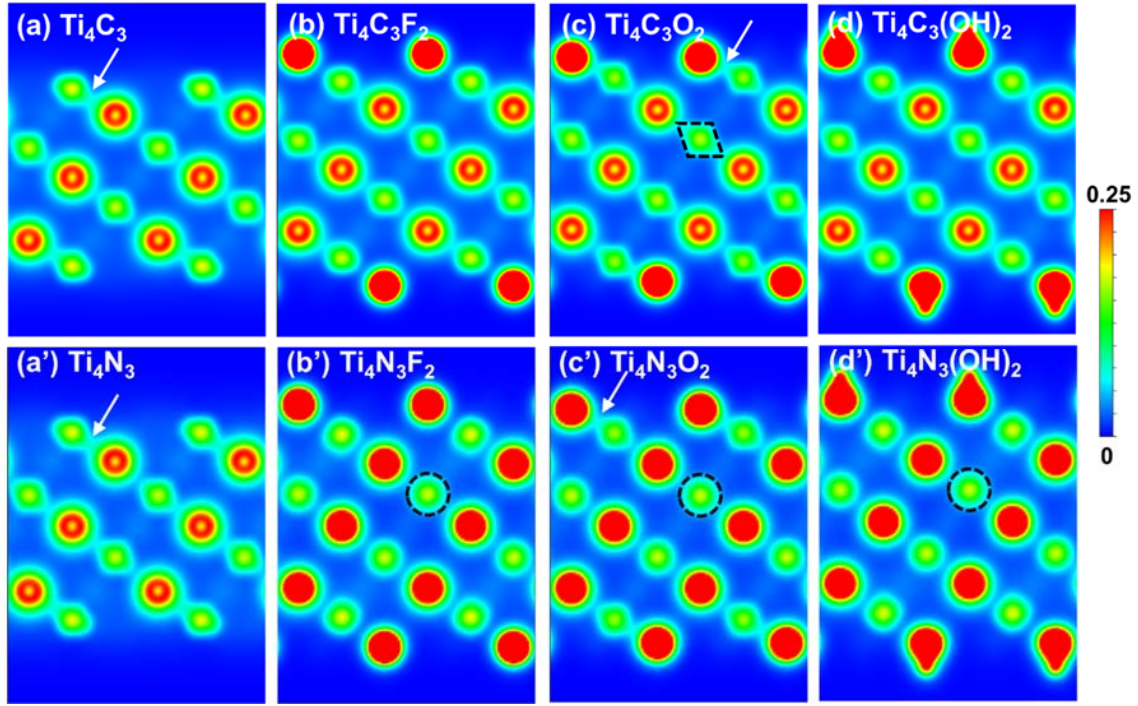

Figure 6. The calculated electron charge density distribution (CDD) for (a-d) $Ti_4C_3$, $Ti_4C_3F_2$, $Ti_4C_3O_2$, and $Ti_4C_3(OH)_2$, and (a'-d') $Ti_4N_3$, $Ti_4N_3F_2$, $Ti_4N_3O_2$, and $Ti_4N_3(OH)_2$ MXenes. The unit of the color bar is e/Å$^3$.

It has been shown that the energy states around the Fermi level are mainly composed of Ti-$d$ orbitals [31]. Therefore, to give a better description of the electronic properties, the total and partial DOS of relevant atomic orbitals in pristine $Ti_{n+1}N_n$/$Ti_{n+1}C_n$ and functionalized $Ti_{n+1}N_nT_2$/$Ti_{n+1}C_nT_2$ are analyzed, as presented in Figure 7 and Figure S6. One can notice that for –F and –OH terminated MXenes, their DOS curves show very similar characters around Fermi level, which are consistent with the observed ELF and CDD behaviors as discussed above. It is well known that a higher DOS at the Fermi level ($N(E_F)$) shows a higher electrical conductivity of materials. To ascertain the effect of terminations on the electrical conductivity, we calculate the values of $N(E_F)$ from Figure 7 and Figure S6, which are listed in Table 3. On one hand, these non-zero $N(E_F)$ indicates that most of the studied MXenes exhibit high metallic electrical conductivity, which is in agreement with the previous studies [46, 57, 58], except that $Ti_2CO_2$ has a zero $N(E_F)$ (Figure S6c and Table 3). This is because each O atom receives two electrons from the surface Ti atom such that the Fermi energy shifts downward to the center of the gap (Figure S6c), and accordingly $Ti_2CO_2$ becomes semiconducting, which is also consistent with the previous studies [18, 31]. On the other hand, the Fermi energy is noted to shift downward and the value of $N(E_F)$ decrease by surface functionalization, and as a result the electrical conductivity is weakened. Furthermore, –O terminated MXenes, particularly $Ti_{n+1}C_nO_2$, show much smaller $N(E_F)$ and consequently exhibit lower electrical conductivity than $Ti_{n+1}C_nF_2$ and $Ti_{n+1}C_n(OH)_2$, which is consistent with the experimental results obtained by Lai et al. [58], regarding the carrier transport behavior of 2D $Ti_2CT_x$.

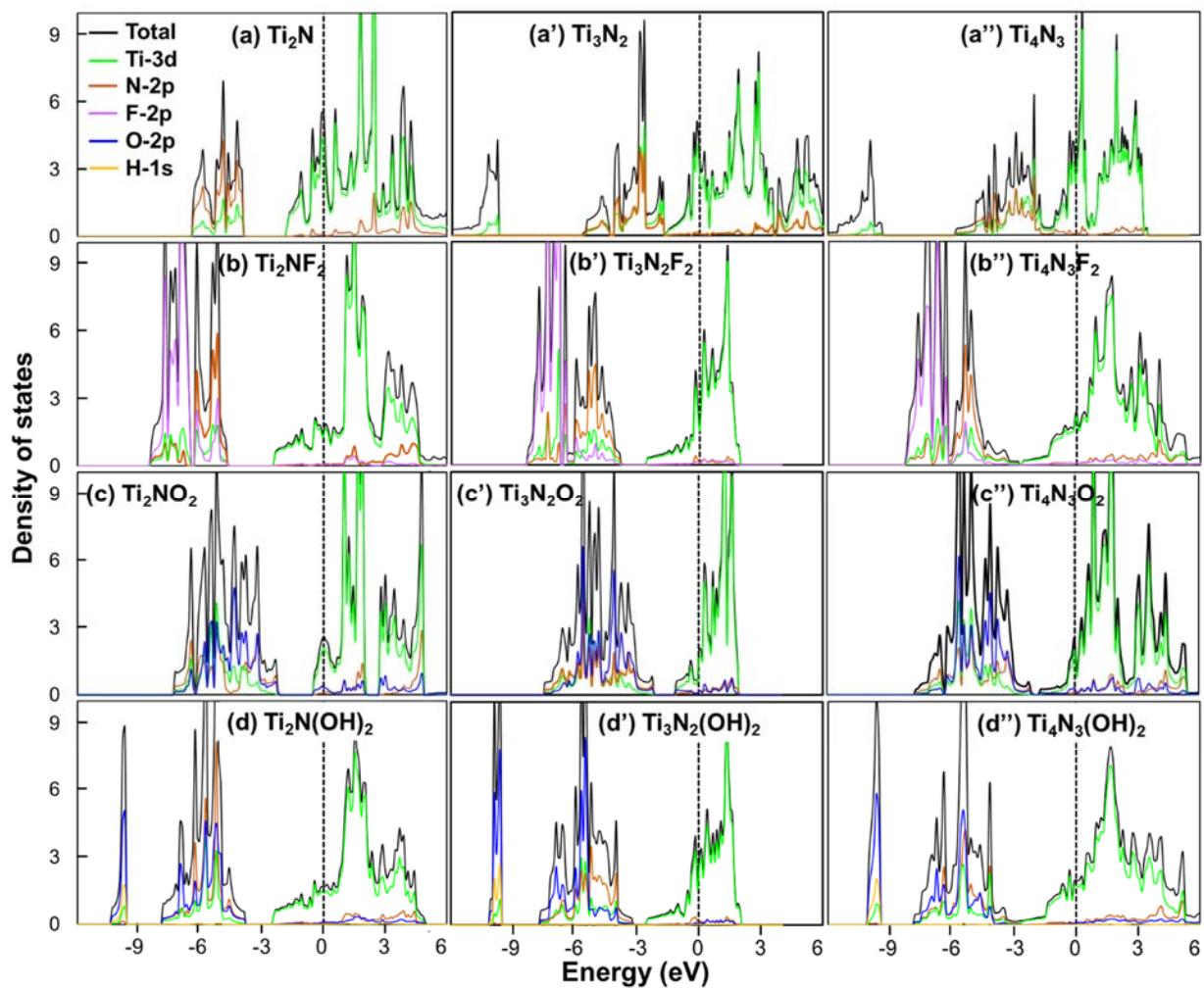

Figure 7. Total and partial densities of states (DOS) of (a-d) $Ti_2N$, $Ti_2NF_2$, $Ti_2NO_2$, $Ti_2N(OH)_2$, (a'-d') $Ti_3N_2$, $Ti_3N_2F_2$, $Ti_3N_2O_2$, $Ti_3N_2(OH)_2$, and (a''-d'') $Ti_4N_3$, $Ti_4N_3F_2$, $Ti_4N_3O_2$, and $Ti_4N_3(OH)_2$ MXenes.

The DOS of pristine $Ti_{n+1}N_n$ shows similar features to those of pristine $Ti_{n+1}C_n$. However, the small valence bands between -9 and -12 eV that are shown in $Ti_{n+1}C_n/Ti_{n+1}C_nT_2$ (Figure S6), $Ti_3N_2$ and $Ti_4N_3$ (Figure 7a', a''), are not observed in $Ti_{n+1}N_nT_2$. Fermi energy downshift behavior after functionalization in $Ti_{n+1}C_nT_2$ is also observed in $Ti_{n+1}N_nT_2$. A much larger energy gap (1.0 – 2.0 eV) shows up between Fermi level and the main valence band in $Ti_{n+1}N_nT_2$. In the case of $Ti_2NO_2$, the Fermi energy has moved pretty close to this gap. Besides, it can be seen from Figure 7 that in $Ti_{n+1}N_nF_2$ and $Ti_{n+1}N_n(OH)_2$ the N-$2p$ orbitals contribute more than Ti-$3d$ to the main valence bands, which is different from the situations in $Ti_{n+1}C_nT_2$ that Ti-$3d$ and C-$2p$ orbitals equally contribute to the overall valence bands (Figure S6). Such an observation indicates that comparing to Ti-C, the ionic bonding interaction between Ti-N becomes stronger and covalent interaction is weakened, because Ti-N-Ti bond is a mixture of metallic, covalent and ionic interactions. Similar to $Ti_{n+1}C_nT_2$, functionalization is found to decrease $N(E_F)$ and therefore shows a negative effect on the electrical conductivity of $Ti_{n+1}N_nT_2$, which is consistent with a previous first principle study [18].

Moreover, it is noted from Table 3 that –F and –OH terminated nitride and carbide MXenes usually exhibit higher $N(E_F)$ than –O terminated ones, except that $Ti_2NO_2$ has a higher $N(E_F)$, make it to have a higher electrical conductivity than $Ti_2NF_2$ and $Ti_2N(OH)_2$. It is also worth emphasizing that the $N(E_F)$ of $Ti_{n+1}N_n$ and $Ti_{n+1}N_nT_2$ are much higher than those of $Ti_{n+1}C_n$ and $Ti_{n+1}C_nT_2$, which suggests that nitride-based MXenes have higher electrical conductivity than carbide-based MXenes. To better understand the surface functionalization effect on electronic properties of nitride/carbide-based MXenes, we analyze the total and partial DOS in Figure 7 and Figure S6 in more detail in the Supporting Information.

Table 3. Evolution of DOS at the Fermi level $N(E_F)$ for $Ti_{n+1}C_nT_2$ and $Ti_{n+1}N_nT_2$ (T = –F, –O, –OH; n = 1, 2, 3) MXenes.

| $Ti_{n+1}C_nT_2$ | $N(E_F)$ | $Ti_{n+1}N_nT_2$ | $N(E_F)$ |
| --- | --- | --- | --- |
| $Ti_2C$ | 4.14 | $Ti_2N$ | 5.50 |
| $Ti_2CF_2$ | 1.62 | $Ti_2NF_2$ | 1.66 |
| $Ti_2CO_2$ | 0 | $Ti_2NO_2$ | 2.52 |
| $Ti_2C(OH)_2$ | 1.34 | $Ti_2N(OH)_2$ | 1.69 |
| $Ti_3C_2$ | 3.48 | $Ti_3N_2$ | 3.90 |
| $Ti_3C_2F_2$ | 1.34 | $Ti_3N_2F_2$ | 2.35 |
| $Ti_3C_2O_2$ | 0.66 | $Ti_3N_2O_2$ | 1.33 |
| $Ti_3C_2(OH)_2$ | 1.16 | $Ti_3N_2(OH)_2$ | 2.47 |
| $Ti_4C_3$ | 4.51 | $Ti_4N_3$ | 3.27 |
| $Ti_4C_3F_2$ | 0.96 | $Ti_4N_3F_2$ | 2.04 |
| $Ti_4C_3O_2$ | 0.63 | $Ti_4N_3O_2$ | 1.77 |
| $Ti_4C_3(OH)_2$ | 1.46 | $Ti_4N_3(OH)_2$ | 1.98 |

## 4. Conclusions

In summary, we have carried out DFT calculations to systematically investigate the structural, elastic and electronic properties of 2D pristine ($Ti_{n+1}C_n/Ti_{n+1}N_n$) and functionalized ($Ti_{n+1}C_nT_2/Ti_{n+1}N_nT_2$) titanium carbide/nitride-based MXenes. Our simulation results show that $Ti_{n+1}C_n/Ti_{n+1}C_nT_2$ have larger lattice constants and monolayer thicknesses than $Ti_{n+1}N_n/Ti_{n+1}N_nT_2$.

The calculated normalized cohesive energy, $\bar{E}_{coh}$, reveals that the structural stability of the studied MXenes increases with increasing the monolayer thickness. The fully functionalized MXenes are more stable than the pristine $Ti_{n+1}C_n/Ti_{n+1}N_n$, especially the –O terminated ones. $Ti_{n+1}C_n$ and $Ti_2N$ prefer to adhere to –O and –F groups rather than –OH group. The nitride-based MXenes are also demonstrated to be less stable with respect to the carbide-based MXenes, as previously reported experiments. However, adsorption energy calculations show that terminal groups prefer to adhere onto the surfaces of $Ti_{n+1}N_n$, and among the studied terminal groups –O is the most favorable group, owing to its relatively low adsorption energy.

In terms of elastic properties, the in-plane Young's moduli of pristine $Ti_{n+1}C_n/Ti_{n+1}N_n$ are noticed to decrease with the increase of monolayer thickness. The overall Young's moduli of $Ti_{n+1}N_n$ are higher than those of $Ti_{n+1}C_n$. Their in-plane Poisson's ratios are larger than those of the corresponding bulk materials.

As for electronic properties, ELF and CDD are used to evaluate the chemical bonding interactions between elements. Although Ti-C/N-Ti is a mixture of metallic, covalent and ionic bonds, ionic bonding is noted to be the dominant interaction between Ti-C/N, as well as between Ti-F/O/OH. The existence of NFE states is also identified in –OH functionalized MXenes. In addition, $Ti_{n+1}N_n(OH)_2$ is found to have more dense NFE states than $Ti_{n+1}C_n(OH)_2$, and therefore has much more potential applications in nanoelectronic devices, catalyst and sensor. Functionalization has a negative effect on the C-C and N-N covalent interaction, but show positive effect on the Ti-N ionic interaction in $Ti_{n+1}N_nT_2$. The original covalent bonding between surface Ti and N atoms completely or partially transfer to the essential ionic interaction after functionalization.

ELF analysis shows the formation of high density lone pair electrons on the surfaces of pristine $Ti_{n+1}C_n/Ti_{n+1}N_n$, which explains their high metallic conductivity predicted by DOS results. DOS analysis also shows that in most cases –O functionalized MXenes exhibit lower electrical conductivity than the –F and –OH functionalized ones. However, $Ti_2NO_2$ has a higher $N(E_F)$ than $Ti_2NF_2$ and $Ti_2N(OH)_2$, and thus has a higher electrical conductivity. Overall, the results show that Nitride-based MXenes have higher electrical conductivities than carbide-based MXenes.

Our DFT calculation results demonstrated that the newly synthesized MXenes, 2D $Ti_{n+1}N_nT_2$ have active surface chemistry, and good elastic and electronic properties despite their low structural stability in etchants. The implications of this work can be helpful to design more MXenes with better performance. Further experimental and computational investigations on $Ti_{n+1}N_nT_2$ MXenes are highly desirable to bring insights into their prospects as advanced materials.

**Acknowledgement**

The authors are grateful for computer time allocation provided by the Extreme Science and Engineering Discovery Environment (XSEDE), award number TG-DMR140008.

**Supporting Information**

# Superior Structural, Elastic and Electronic Properties of 2D Titanium Nitride MXenes Over Carbide MXenes: A Comprehensive First Principles Study

Ning Zhang[1], Yu Hong[1], Sanaz Yazdanparast[1], and Mohsen Asle Zaeem[1,2*]

[1] Department of Materials Science and Engineering, Missouri University of Science and Technology, Rolla, MO 65409, USA

[2] Department of Mechanical Engineering, Colorado School of Mines, Golden, CO 40801, USA

*Corresponding author; email: zaeem@mines.edu; zaeem@mst.edu (M. Asle Zaeem).

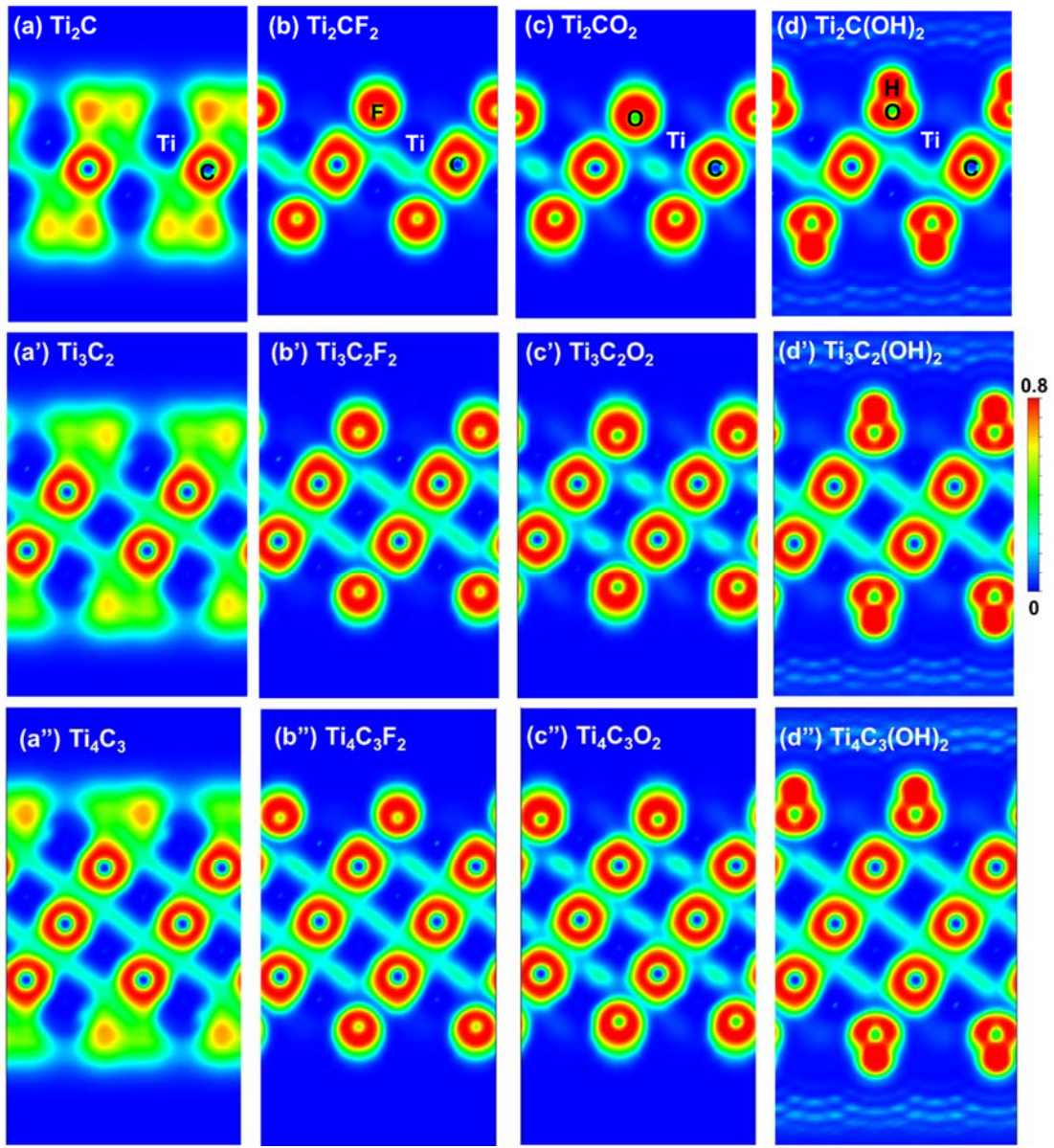

Figure S1. The ELF contour plots projected on the $(1\bar{1}0)$ plane of 2D (a-d) $Ti_2C$, $Ti_2CF_2$, $Ti_2CO_2$, $Ti_2C(OH)_2$, (e-h) $Ti_3C_2$, $Ti_3C_2F_2$, $Ti_3C_2O_2$, $Ti_3C_2(OH)_2$, and (i-l) $Ti_4C_3$, $Ti_4C_3F_2$, $Ti_4C_3O_2$, $Ti_4C_3(OH)_2$.

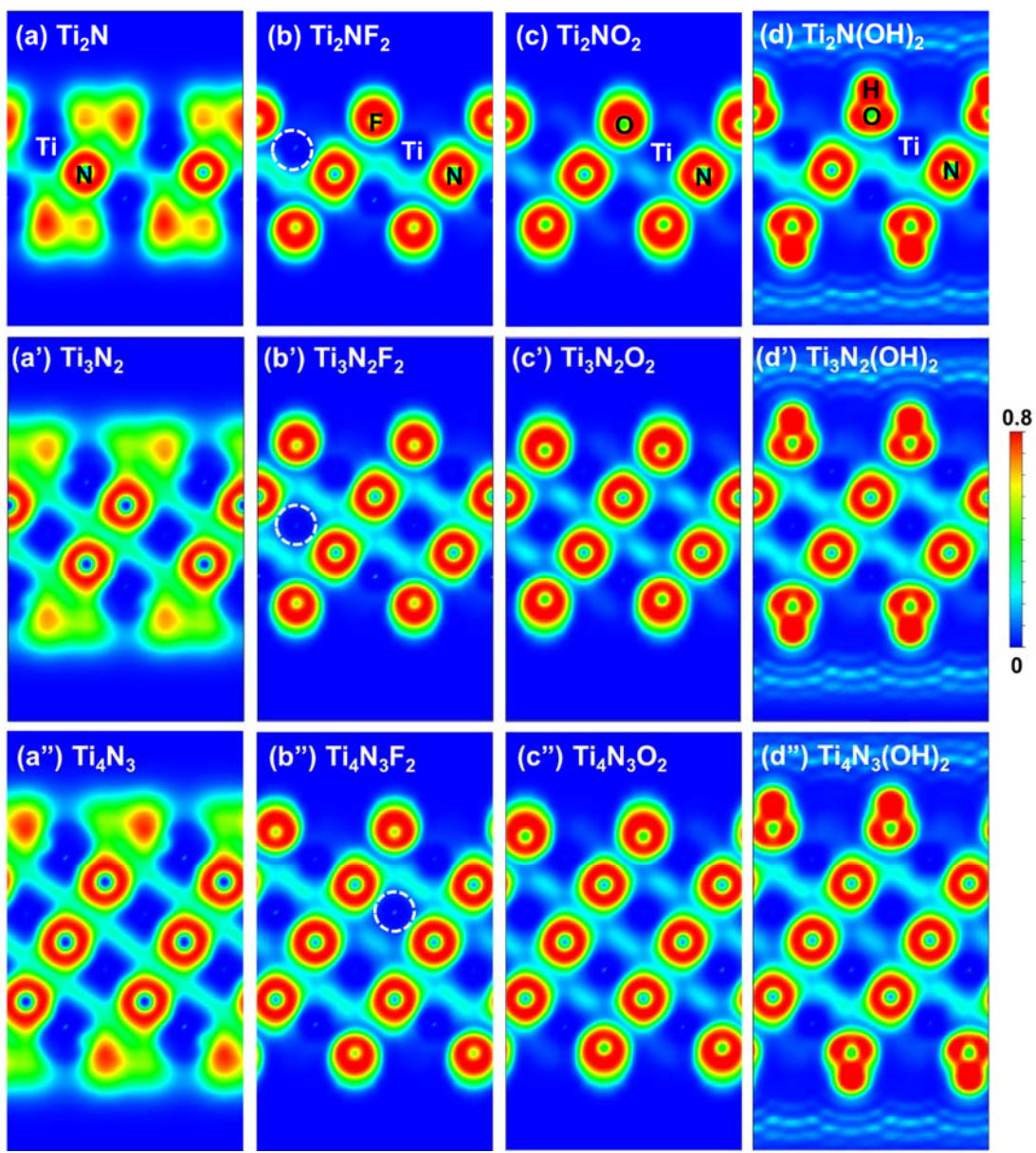

Figure S2. The ELF contour plots projected on the $(1\bar{1}0)$ plane of 2D (a-d) $Ti_2N$, $Ti_2NF_2$, $Ti_2NO_2$, $Ti_2N(OH)_2$, (e-h) $Ti_3N_2$, $Ti_3N_2F_2$, $Ti_3N_2O_2$, $Ti_3N_2(OH)_2$, and (i-l) $Ti_4N_3$, $Ti_4N_3F_2$, $Ti_4N_3O_2$, $Ti_4N_3(OH)_2$.

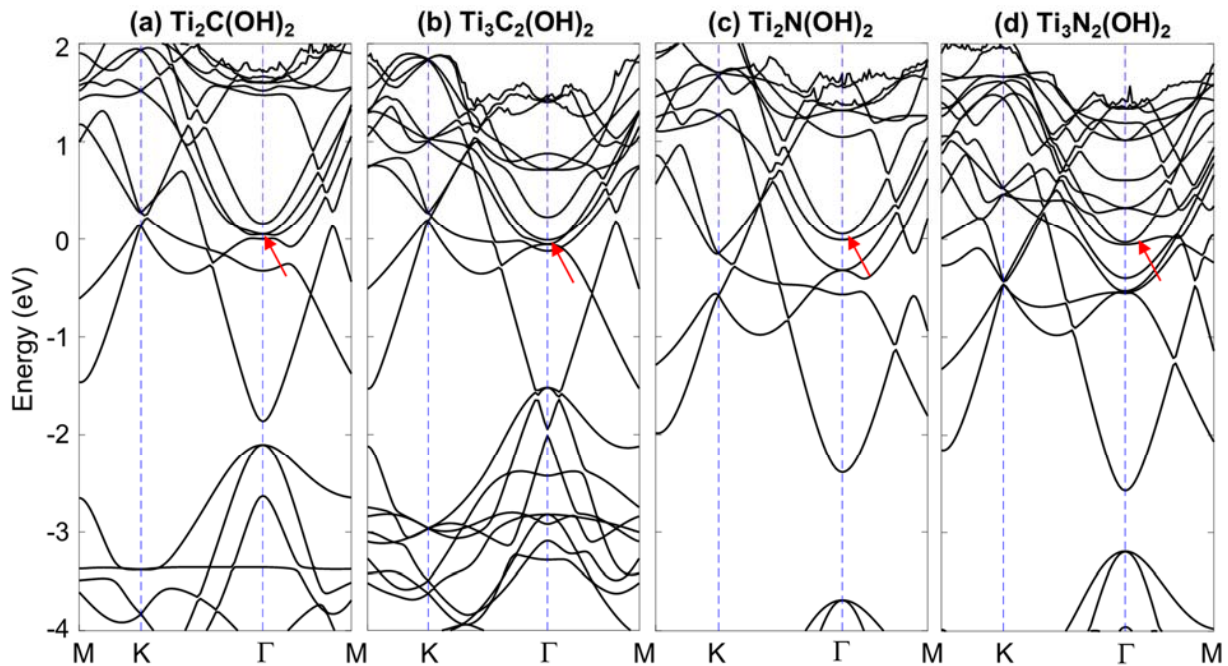

Figure S3. Projected energy band structures of –OH terminated MXenes: (a) $Ti_2C(OH)_2$, (b) $Ti_3C_2(OH)_2$, (c) $TiN_2(OH)_2$ and (d) $Ti_3N_2(OH)_2$. The red arrows indicate the lowest energy NFE near the Fermi energy at the Γ point.

It is evident that the highest CDD resides in the immediate vicinity of the non-metal ions nuclei. Comparing Figure S4a, a' and a'', it is noted that with the increase of the monolayer thickness the interaction between C atoms and the surface Ti atoms in pristine $Ti_{n+1}C_n$ becomes stronger, as indicated by white arrows (Figure S4a, a', a''), where the density of overlapped electron cloud in $Ti_4C_3$ (Figure S4a'') is the highest. For the cases of $Ti_{n+1}C_nF_2$ and $Ti_{n+1}C_n(OH)_2$ MXenes, with the increase of monolayer thickness the shape of CDD around Ti nuclei changes from its original perfect sphere to extended sphere along Ti-F, Ti-C and Ti-(OH) bond path. However, since overlapping electron cloud is not observed, the dominant interaction between Ti-F and Ti-OH is still ionic bonding. In $Ti_{n+1}C_nO_2$, the electron cloud outside the Ti nucleus is significantly stretched along the Ti-O and Ti-C bond path, which results in an approximate rhombus geometry, as denoted in Figure S4. Such elongation results in slight overlaps of electron cloud between Ti-O and Ti-C, which indicates that a small partial of ionic bonding transfers to covalent bonding through sharing electrons. And also it demonstrates that the terminal groups are more likely to bond with the surface Ti atoms as the electron donor. Although Ti-C bond is essentially ionic in all studied MXenes, the density of overlapping electron pockets along the C-C bond path in Figure S4 implies that the C-C covalent bonds in pristine $Ti_{n+1}C_n$, $Ti_{n+1}C_nF_2$ and $Ti_{n+1}C_n(OH)_2$ are stronger than those in $Ti_{n+1}C_nO_2$, despite that this covalent bonding is relatively weak. Therefore, we can conclude that functionalization has a negative effect on the C-C covalent interaction, the strength of which is in the order of: $Ti_{n+1}C_n > Ti_{n+1}C_nF_2/Ti_{n+1}C_n(OH)_2 > Ti_{n+1}C_nO_2$.

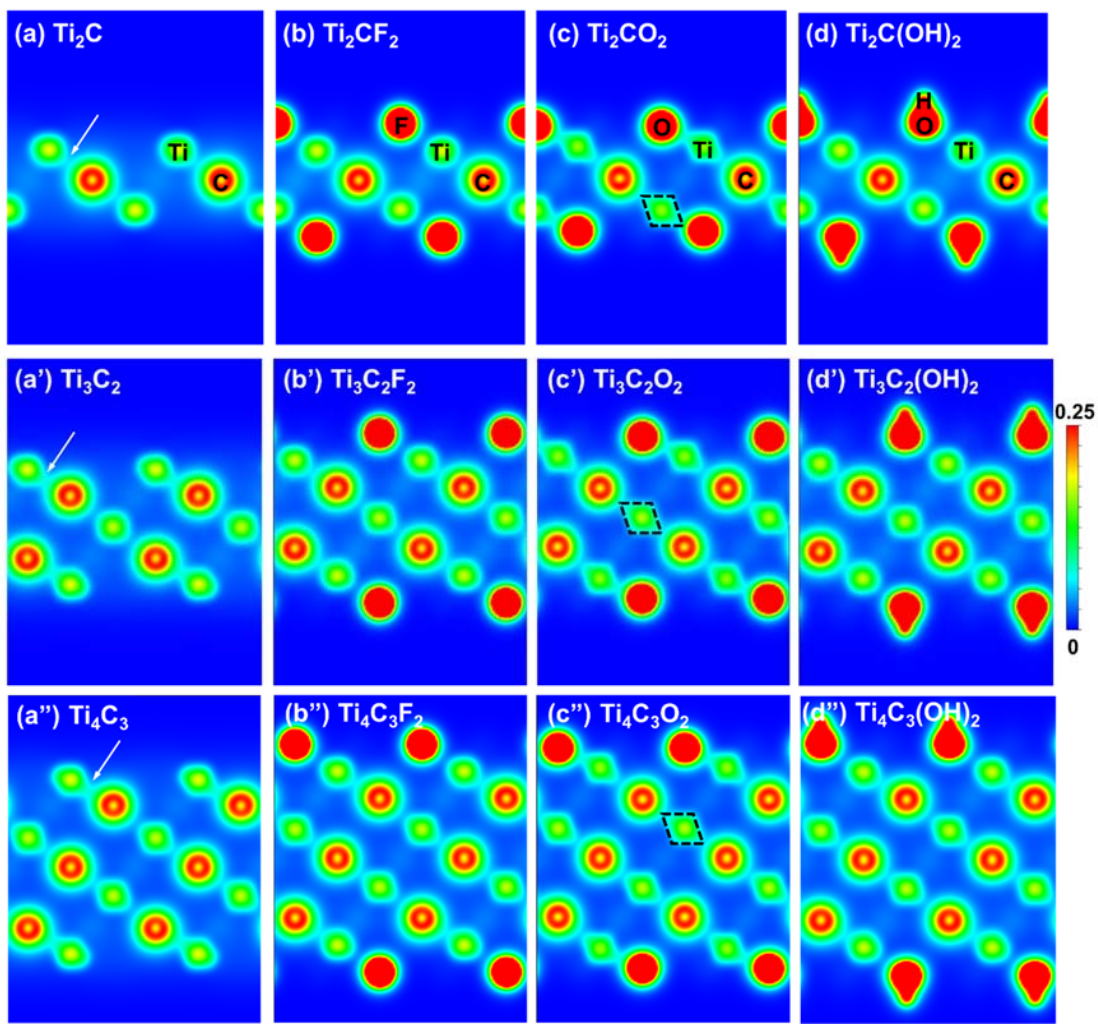

Figure S4. The calculated electron charge density distribution (CDD) projected on the $(1\bar{1}0)$ plane of 2D (a-d) $Ti_2C$, $Ti_2CF_2$, $Ti_2CO_2$, $Ti_2C(OH)_2$, (e-h) $Ti_3C_2$, $Ti_3C_2F_2$, $Ti_3C_2O_2$, $Ti_3C_2(OH)_2$, and (i-l) $Ti_4C_3$, $Ti_4C_3F_2$, $Ti_4C_3O_2$, $Ti_4C_3(OH)_2$.

The surface effect on CDD around Ti atoms becomes more apparent in $Ti_{n+1}N_nO_2$. The elongated electron cloud of Ti atoms along Ti-N bond path in pristine $Ti_{n+1}N_n$ is receded when –O groups are added (Figure S5c, c', c'').

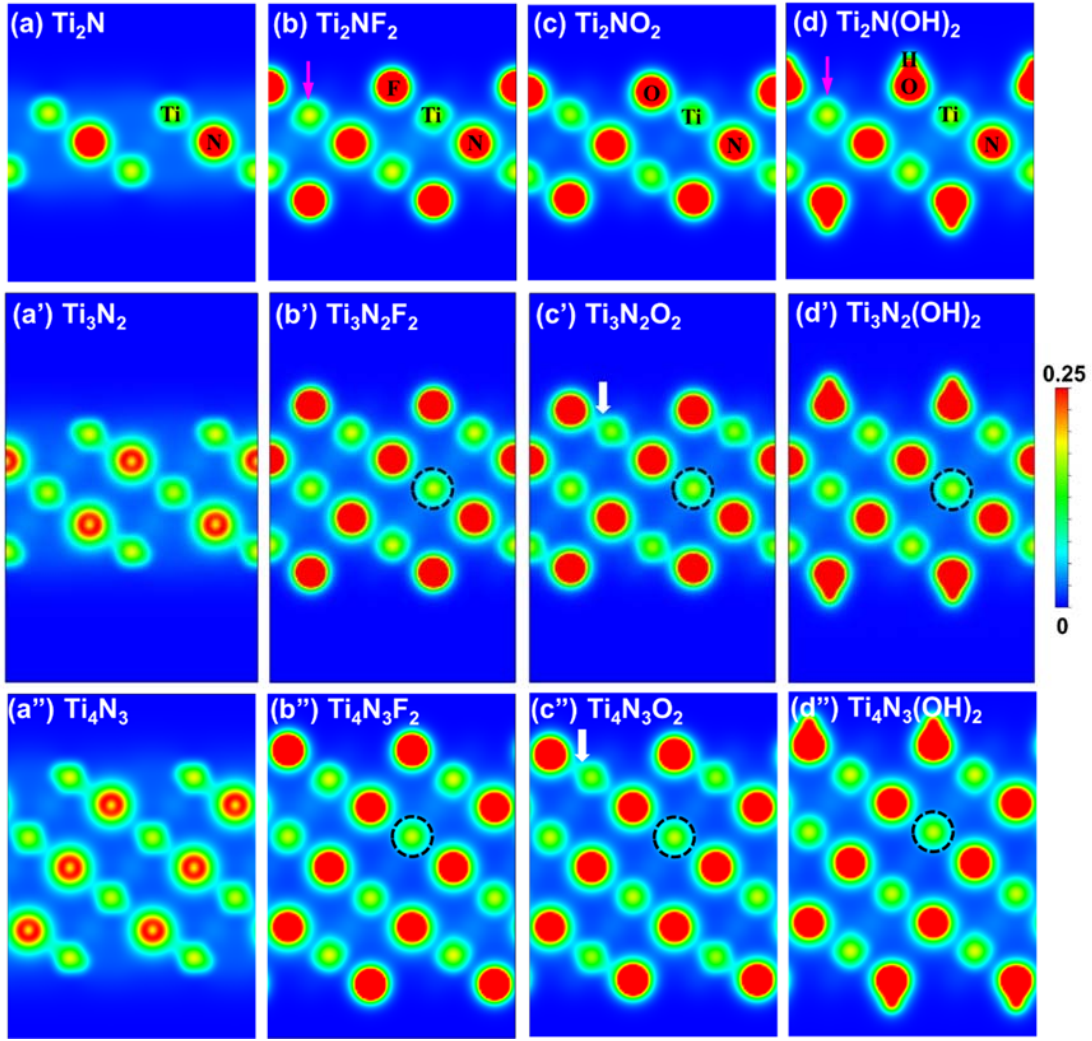

Figure S5. The calculated electron charge density distribution (CDD) projected on the (1$\bar{1}$0) plane of 2D (a-d) Ti$_2$N, Ti$_2$NF$_2$, Ti$_2$NO$_2$, Ti$_2$N(OH)$_2$, (e-h) Ti$_3$N$_2$, Ti$_3$N$_2$F$_2$, Ti$_3$N$_2$O$_2$, Ti$_3$N$_2$(OH)$_2$, and (i-l) Ti$_4$N$_3$, Ti$_4$N$_3$F$_2$, Ti$_4$N$_3$O$_2$, Ti$_4$N$_3$(OH)$_2$.

For the carbide-based MXenes, one can observe that in all the cases the DOS bands around Fermi level are primarily contributed by Ti-3$d$. The conduction states (the band above Fermi level) are mainly composed of Ti-3$d$. In contrast, the valence bands between -3.0 and -6.0 eV (Figure S6a) are primarily made of C-2$p$ in pristine Ti$_2$C, but are formed equally by the hybridized Ti-3$d$ and C-2*p* orbitals in the case of Ti$_3$C$_2$ and Ti$_4$C$_3$. Also, as aforementioned by examining the band structures near the Fermi level we find that the Fermi energy shifts downward to the lower states as after functionalization. In other words, a bunch of valence bands transfer to the conduction states. This is due to the strong hybridization between orbitals of the terminal groups and the surface Ti atoms. After –F functionalization, new states are formed below the Fermi energy (around -6.0 eV), and the new bands are mainly made of the F-2*p* orbitals, as can be seen in Figure S6b, b', b''. Similar phenomenon are also observed in –O and –OH functionalized MXenes, whereas the O-2*p* orbitals dominate the new formed bands (Figure S6c, c', c'' and Figure S6d, d', d''). For the case of –OH functionalized MXenes, additional sharp and narrow bands, which are contributed mainly by O-2*p* orbitals, are also observed at around -9.0 eV (Figure S6d, d', d'').

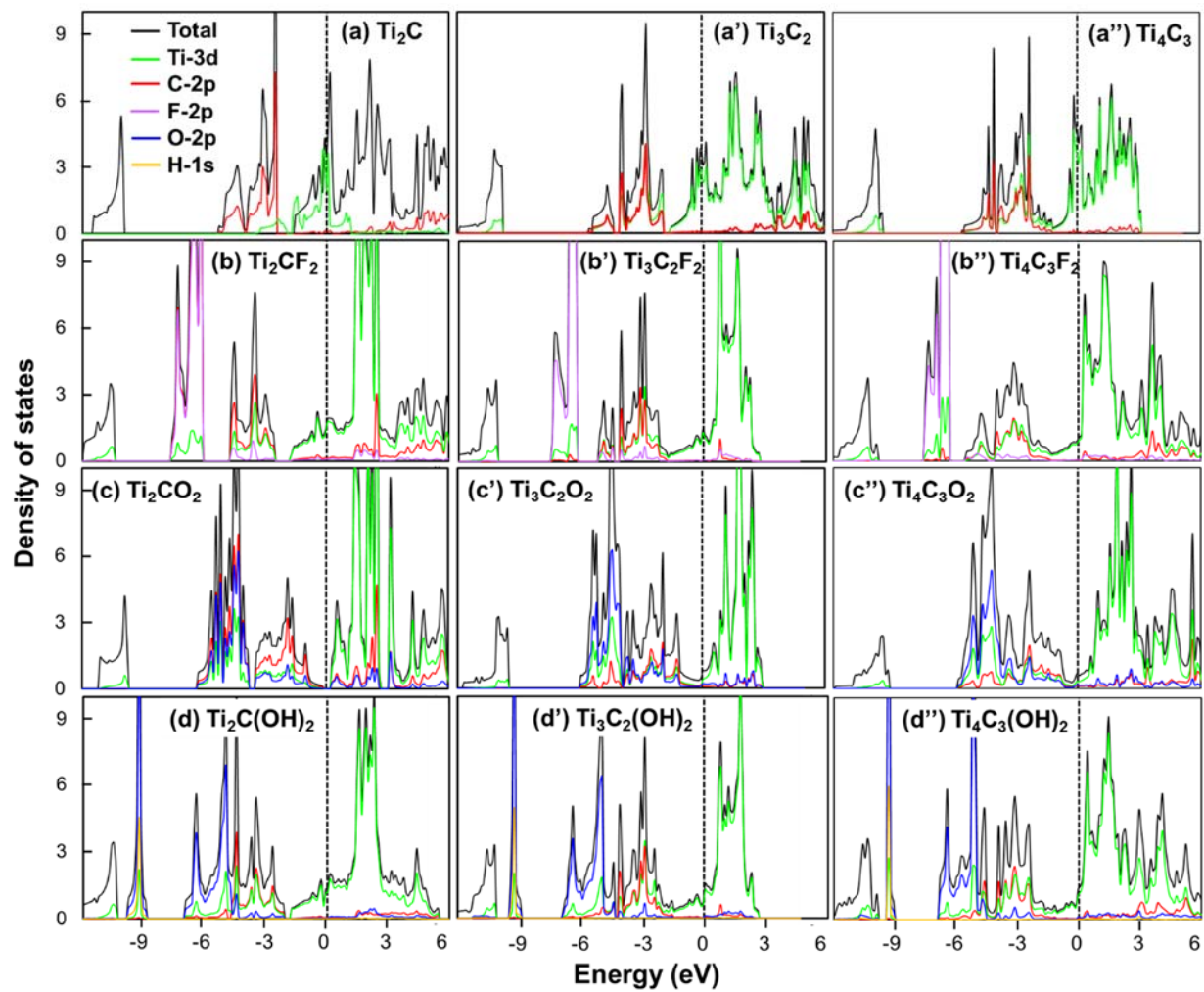

Figure S6. Total and partial DOS of (a-d) $Ti_2C$, $Ti_2CF_2$, $Ti_2CO_2$, $Ti_2C(OH)_2$, (a'-d') $Ti_3C_2$, $Ti_3C_2F_2$, $Ti_3C_2O_2$, $Ti_3C_2(OH)_2$, and (a''-d'') $Ti_4C_3$, $Ti_4C_3F_2$, $Ti_4C_3O_2$, $Ti_4C_3(OH)_2$.

Table S1. Calculated normalized cohesive energy $\bar{E}_{coh}$ (eV/atom) for pristine $Ti_{n+1}C_n/Ti_{n+1}N_n$ and functionalized $Ti_{n+1}C_nT_2/Ti_{n+1}N_nT_2$ (T = –F, –O, –OH; n = 1, 2, 3) monolayers, in comparison with available first-principles calculations from the literature.

| $Ti_{n+1}C_nT_2$ | $\bar{E}_{coh}$ (eV/atom) | $Ti_{n+1}N_nT_2$ | $\bar{E}_{coh}$ (eV/atom) |
|---|---|---|---|
| $Ti_2C$ | -6.12 (-6.28 [1], -6.64 [2]) | $Ti_2N$ | -5.81 (-5.95 [1]) |
| $Ti_2CF_2$ | -6.36 | $Ti_2NF_2$ | -6.11 |
| $Ti_2CO_2$ | -6.96 | $Ti_2NO_2$ | -6.81 |
| $Ti_2C(OH)_2$ | -6.13 | $Ti_2N(OH)_2$ | -5.85 |
| $Ti_3C_2$ | -6.54 (-6.82 [1]) | $Ti_3N_2$ | -5.95 (-6.38 [1]) |
| $Ti_3C_2F_2$ | -6.69 | $Ti_3N_2F_2$ | -6.31 |
| $Ti_3C_2O_2$ | -7.05 | $Ti_3N_2O_2$ | -6.85 |
| $Ti_3C_2(OH)_2$ | -6.55 | $Ti_3N_2(OH)_2$ | -5.99 |
| $Ti_4C_3$ | -6.73 (-7.0 [1]) | $Ti_4N_3$ | -6.06 (-6.6 [1]) |
| $Ti_4C_3F_2$ | -6.83 | $Ti_4N_3F_2$ | -6.47 |
| $Ti_4C_3O_2$ | -7.11 | $Ti_4N_3O_2$ | -6.87 |
| $Ti_4C_3(OH)_2$ | -6.73 | $Ti_4N_3(OH)_2$ | -6.21 |

Table S2. The adsorption energies ($E_{ads}$) for adhesion of terminal groups (-F, -O, and -OH) onto the surfaces of pristine $Ti_{n+1}C_n$/$Ti_{n+1}N_n$ (n = 1, 2, 3), in comparison with available first-principles calculations from the literature (inside the parentheses). Energy unit is eV/atom or eV/molecule for –OH.

|  | $Ti_{n+1}X_nF_2$ | $Ti_{n+1}X_nO_2$ | $Ti_{n+1}X_n(OH)_2$ |
| --- | --- | --- | --- |
| X=C | -6.916 (-6.358 [3]) | -8.923 (-7.057 [3]) | -5.664 (-5.077 [3]) |
| X=N | -7.451 | -9.993 | -6.268 |